\documentclass[sigconf]{acmart}
\acmConference[MSR 2026]{MSR '26: Proceedings of the 23rd International Conference on Mining Software Repositories}{April 2026}{Rio de Janeiro, Brazil}

\newif\ifDEBUG
\DEBUGtrue

\newif\ifEXTENDED
\EXTENDEDtrue
\EXTENDEDfalse

\usepackage{graphicx}
\usepackage{hyperref} 
\usepackage{cleveref}
\usepackage{caption}
\usepackage{pifont}
\usepackage{xcolor}
\usepackage{listings}
\usepackage{minted}
\usepackage{subcaption}
\usepackage{caption}

\usepackage{booktabs}
\usepackage{multirow} 
\usepackage{subcaption}
\usepackage{makecell}
\usepackage{tabularx}
\usepackage{ragged2e}
\usepackage{xspace}
\usepackage{soul}
\usepackage{array} 
\usepackage{float}

\usepackage{amsmath}
\usepackage{textcomp}
\usepackage{colortbl}
\usepackage{longtable}
\usepackage{array}
\usepackage{adjustbox}
\usepackage{algorithm}
\usepackage{threeparttable}
\usepackage{tablefootnote}
\usepackage{booktabs} 
\usepackage{amsmath}
\usepackage{algorithm}
\usepackage{algpseudocode}
\usepackage[normalem]{ulem} 
\usepackage{xspace}
\usepackage{relsize}
\usepackage{multirow} 
\usepackage{balance} 
\usepackage{fancyvrb} 
\usepackage{caption}
\usepackage{pifont}

\usepackage{enumitem}
\setlist[itemize]{leftmargin=*,noitemsep,topsep=0pt}
\setlist[enumerate]{leftmargin=*}

\usepackage{etoolbox}
\makeatletter
\patchcmd{\@makecaption}
	{\scshape}
	{}
	{}
	{}
\makeatletter
\patchcmd{\@makecaption}
	{\\}
	{.\ }
	{}
	{}
\makeatother

\newcommand{\ie}{\textit{i.e.,}\xspace}
\newcommand{\eg}{\textit{e.g.,}\xspace}
\newcommand{\etal}{\textit{et al.}\xspace}

\usepackage{mdframed}
\mdfsetup{skipabove=0.5\topskip,skipbelow=0.5\topskip,align=center}

\usepackage{amsthm}
\newtheorem{thm}{Theorem}

\DeclareMathSymbol{\mlq}{\mathord}{operators}{``}
\DeclareMathSymbol{\mrq}{\mathord}{operators}{`'}

\newif\ifSAVESPACE
\SAVESPACEfalse

\ifSAVESPACE

\else

\fi

\usepackage{todonotes}
\usepackage{soul}
\usepackage{fontawesome}
\usepackage{ragged2e}

\ifDEBUG
    \newcommand{\AH}[1]{\todo[color=cyan,inline]{AH:#1}}
    \newcommand{\AM}[1]{\todo[color=red,inline]{Machiry:#1}}
    \newcommand{\JD}[1]{\todo[color=yellow,inline]{JD:#1}}
    \newcommand{\SA}[1]{\todo[color=green,inline]{SA:#1}}
    \newcommand{\PA}[1]{\todo[color=orange,inline]{PA:#1}}
    
    \newcommand{\KR}[1]{\todo[color=yellow,inline]{Kyle:#1}}
    \newcommand{\LS}[1]{\todo[color=green,inline]{LS:#1}}
    \newcommand{\HP}[1]{\todo[color=cyan,inline]{HP:#1}}
    \newcommand{\NJE}[1]{\todo[color=red,inline]{NJE: #1}}
    \newcommand{\GKT}[1]{\todo[color=red,inline]{GKT:#1}}
    
    \newcommand{\RH}[1]{\todo[color=red,inline]{RH:#1}}
    \newcommand{\WJ}[1]{\todo[color=SkyBlue,inline]{Wenxin:#1}} 
    \newcommand{\KC}[1]{\todo[color=orange,inline]{Kelechi Says:#1}}
    \newcommand{\AG}[1]{\todo[color=orange,inline]{AG:#1}}
    \newcommand{\PJ}[1]{\todo[color=lime,inline]{PJ:#1}}
    \newcommand{\AZ}[1]{\todo[color=teal,inline]{Antonio:#1}}
    \newcommand{\PT}[1]{\todo[color=pink,inline]{Parth:#1}}
    
\else
    \newcommand{\AH}[1]{}
    \newcommand{\AM}[1]{}
    \newcommand{\JD}[1]{}
    \newcommand{\SA}[1]{}
    \newcommand{\PA}[1]{}
    \newcommand{\KR}[1]{}
    \newcommand{\LS}[1]{}
    \newcommand{\HP}[1]{}
    \newcommand{\NJE}[1]{}
    \newcommand{\GKT}[1]{}
    \newcommand{\KC}[1]{}
    \newcommand{\RH}[1]{}
    \newcommand{\WJ}[1]{}
    \newcommand{\AG}[1]{}
    \newcommand{\PJ}[1]{}
    \newcommand{\PT}[1]{}
    \newcommand{\AZ}[1]{}
    
\fi

\usepackage{hyperref}

\usepackage{cleveref}
\crefformat{section}{\S#2#1#3}
\crefname{figure}{Figure}{Figures}
\crefname{table}{Table}{Tables}
\crefname{theorem}{Theorem}{Theorems}
\crefname{thm}{Theorem}{Theorems}
\crefname{lemma}{Lemma}{Lemmata}
\crefname{equation}{Eqt.}{Eqts.}
\crefformat{Grammar}{Grammar #1}
\crefname{appendix}{Appendix}{Appendices}
\crefname{listing}{Listing}{Listings}

\usepackage{url}

\usepackage{tcolorbox}

\makeatletter
\newcommand{\linebreakand}{%
  \end{@IEEEauthorhalign}
  \hfill\mbox{}\par
  \mbox{}\hfill\begin{@IEEEauthorhalign}
}
\makeatother

\usepackage{pifont}

 \usepackage{pifont}

\begin{document}

\title{How Do Agents Perform Code Optimization? An Empirical Study}

\author{Huiyun Peng}
\affiliation{%
  \institution{Purdue University}
  \city{West Lafayette}
  \state{Indiana}
  \country{USA}
}

\author{Antonio Zhong}
\affiliation{%
  \institution{Purdue University}
  \city{West Lafayette}
  \state{Indiana}
  \country{USA}
}

\author{Ricardo Andrés Calvo Méndez}
\affiliation{%
  \institution{Purdue University}
  \city{West Lafayette}
  \state{Indiana}
  \country{USA}
}

\author{Kelechi G. Kalu}
\affiliation{%
  \institution{Purdue University}
  \city{West Lafayette}
  \state{Indiana}
  \country{USA}
}

\author{James C. Davis}
\affiliation{%
  \institution{Purdue University}
  \city{West Lafayette}
  \state{Indiana}
  \country{USA}
}

\renewcommand{\shortauthors}{Peng et al.}

\begin{abstract}
Performance optimization is a critical yet challenging aspect of software development, often requiring a deep understanding of system behavior, algorithmic tradeoffs, and careful code modifications.
Although recent advances in AI coding agents have accelerated code generation and bug fixing, little is known about how these agents perform on real-world performance optimization tasks.

We present the first empirical study comparing agent- and human-authored performance optimization commits, analyzing 324 agent-generated and 83 human-authored PRs from the AIDev dataset across adoption, maintainability, optimization patterns, and validation practices. 
We find that AI-authored performance PRs are less likely to include explicit performance validation than human-authored PRs (45.7\% vs. 63.6\%, $p=0.007$).
In addition, AI-authored PRs largely use the same optimization patterns as humans.
We further discuss limitations and opportunities for advancing agentic code optimization.
\end{abstract}



\maketitle

\section{Introduction}


Software performance is central to the reliability~\cite{Jain1991}, user experience~\cite{Wilke2013}, and energy efficiency~\cite{krasner2021cost} of modern systems. Despite decades of advances in compilers~\cite{Muchnick1998}, algorithms~\cite{Bader2002}, and hardware-aware tuning~\cite{Buchty2012}, performance engineering remains a labor-intensive task requiring specialized expertise and systematic profiling~\cite{balaprakash2018autotuning}.
Large Language Models (LLMs) are increasingly used for software development tasks such as code synthesis, debugging, and refactoring~\cite{Ozkaya2023}, and have shown promise in improving non-functional properties like runtime and energy efficiency~\cite{gong2025, peng2024largelanguagemodelsenergyefficient}. 

Autonomous coding agents are increasingly contributing substantial numbers of PRs across open-source software, including performance-related changes~\cite{watanabe2025use}.
Performance optimization, however, poses distinct challenges: performance-oriented PRs often have non-local effects, rely on implicit assumptions about workloads or hardware, and require strong empirical evidence (\eg benchmarking or profiling) to justify their correctness and benefit~\cite{goedecker2001performance}.
As AI agents begin to author such changes at scale, it becomes critical to understand how performance improvements are carried out by humans and AI in real-world settings.


To address this gap, we conduct the first empirical study of performance-focused pull requests drawn from the AIDev~\cite{li2025aidev} dataset, comparing agent and human-authored performance PRs via the SysLLMatic~\cite{peng2025sysllmaticlargelanguagemodels} catalog. 
We ask:
\begin{itemize}
    \item \textbf{RQ1:} How do AI agents and humans differ in the optimization patterns they apply in performance-oriented patches?
    \item \textbf{RQ2:} How do AI agents and humans differ in their testing and validation of performance improvements?
\end{itemize}

We answer these questions using a subset of 407 performance-related PRs (4,954 commit details) from the AIDev dataset.
We find that AI-authored performance PRs rely on optimization patterns similar to those used by humans, but offer weaker validation of performance impact and mostly rely on static reasoning rather than benchmarking.
In addition, we contribute an extended performance optimization catalog, introducing 14 new patterns spanning two additional categories to better capture the diversity of real-world performance patches.
This work advances our understanding of real-world performance optimization practices and highlights both the strengths and limitations of AI agents in this domain. 

\section{Background \& Related Work}

\subsection{Performance Engineering}
\subsubsection{Optimization Patterns}
\label{sec:background-pattern}
Performance engineering spans the full software stack, from application logic to operating system and hardware behavior. At the source-code level, performance tuning involves revising algorithms and data structures, improving control flow and memory locality, and exploiting parallelism --- often guided by a system-wide view of where time and resources are spent~\cite{Gregg_2021}. 

To answer RQ1, we require a concrete catalog of such optimization methods so we can consistently characterize the strategies used in performance-related PRs. 
SysLLMatic~\cite{peng2025sysllmaticlargelanguagemodels} provides this foundation by introducing the state-of-the-art catalog of 43 optimization patterns across seven categories, and demonstrating how the catalog can be applied to guide optimizations on real-world benchmarks.
We adopt this catalog as a shared vocabulary for describing performance improvements in our dataset.

\subsubsection{Performance Validation}
\label{sec:background-validation}
Optimizations require rigorous validation because performance is highly sensitive to measurement noise, workload choice, and experimental design; best practices therefore emphasize representative workloads, repeatable protocols, and statistically grounded comparisons~\cite{Jain1991}. 

To answer RQ2, we require a characterization of performance validation practices so we can systematically assess how performance claims are supported in \texttt{perf} PRs.
We focus on three common forms of performance validation evidence used in software engineering practice.
First, \emph{Benchmark-Based Validation} relies on quantitative measurements (\eg unit tests or microbenchmarks) to directly assess performance impact~\cite{performanceunittest2015, Jain1991}. 
Second, \emph{Profiling-Based Validation} uses profiling artifacts such as hotspot traces, CPU samples, or flame graphs to localize performance costs~\cite{gregg2016flame}. 
Third, \emph{Static-Reasoning-Based Validation} supports performance claims through algorithmic arguments without relying on runtime data, \eg by identifying asymptotic inefficiencies~\cite{Olivo2015, Banerjee2014}.
These three forms capture the primary ways developers justify performance improvements.

\subsection{Related Work}
Prior works have mined real-world patches to study performance optimization. RAPGen~\cite{garg2025rapgenapproachfixingcode} mines performance-related PRs to build an optimization knowledge base for fixing inefficiencies, 
while Yi \etal~\cite{yi2025experimentalstudyreallifellmproposed} uses human-authored performance PRs as ground truth, prompt LLMs to optimize the original code, and compare the results against the developers’ patches.
Concurrently, recent work has begun empirically examining agent PRs in real-world repositories, comparing their characteristics and integration outcomes with human-authored PRs~\cite{watanabe2025use, Tufano2024, Jamil2025}.
However, prior work has not compared agent- and human-authored performance pull requests or analyzed performance factors such as optimization strategies and validation practices.
To fill this gap, we present the first empirical comparison of agent- and human-authored performance PRs, analyzing
optimization patterns and validation practices.

\section{Methodology}
\begin{figure}
    \centering
    \includegraphics[width=\linewidth, trim={3.8cm 19.3cm 4.6cm 2cm}, clip]{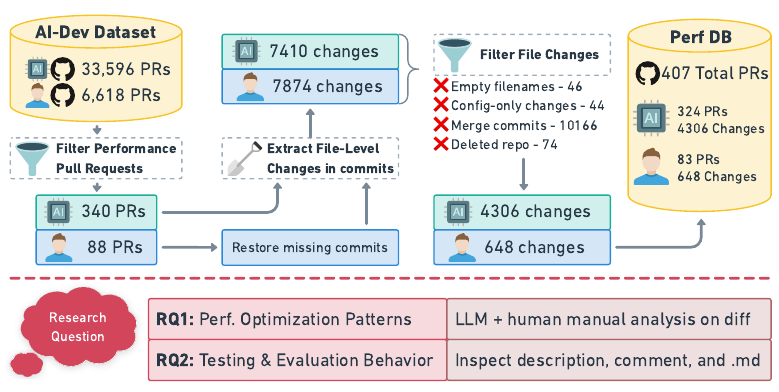}
    \caption{
        Methodology overview illustrating the data collection pipeline and how the resulting dataset supports answering RQ1 and RQ2.
    }
    \vspace{-1.5em}
    \label{fig:methodology-overview}
\end{figure}

This section describes our data collection pipeline and analysis methods for performance-related PRs, as summarized in~\cref{fig:methodology-overview}. 

\subsection{Data Preparation}
We conduct our study using \textsc{AIDev}~\cite{li2025aidev}, a dataset of GitHub pull requests (PRs) authored by agents and humans with commit, diff, review, and repository-level metadata. 
We use two subsets: \textsc{AIDev-Pop} (33{,}596 PRs; repositories with $\ge$100 stars), and \textsc{Human-PR} (6{,}618 PRs; repositories with $\ge$500 stars).
We focus on performance-related PRs (label: \texttt{perf}), yielding 340 agent PRs and 88 human PRs. 

To enable code-level analysis, we require fine-grained diff data.
Because \textsc{Human-PR} omits commit-level diff data, we mine commit details from GitHub and reconstruct 7{,}874 commit-detail records for the 88 human \texttt{perf} PRs; combined with \textsc{AIDev-Pop}, this yields 15{,}284 commit-detail records across the 428 \texttt{perf} PRs. 
We then apply quality filters, removing records with missing filenames~(46), configuration-only changes~(44), merge commits (10{,}166), and commits from deleted repositories~(74), resulting in 4{,}954 valid commit-detail records corresponding to 407 \texttt{perf} PRs with analyzable patches. Coverage remains high, with valid commits in~324/340 AI-agent PRs (95.3\%) and~83/88 human PRs (94.3\%).

\subsection{Optimization Patch Characteristics (RQ1)}

This analysis characterizes the optimization patterns applied in \texttt{perf} PRs authored by AI agents and humans. 
To enable systematic labeling, we adopt the catalog from SysLLMatic~\cite{peng2025sysllmaticlargelanguagemodels} (\cref{sec:background-pattern}).
This catalog organizes performance improvements at two levels of granularity:
high-level categories for broad optimization strategies (\eg Memory/Data Locality Optimization), and finer-grained patterns for specific implementation mechanisms (\eg Caching).

\subsubsection{General Labeling Procedure}
We use LLMs as annotation aids and calibrate via iterative human inspection~\cite{wang2024human}.
We use two commercial LLMs (SOTA as of December 2025): \textsc{GPT-5.1} and \textsc{Gemini-3-Pro-Preview}, with temperature=0, to independently classify each patch.
When the models agree, we accept
the label, and when they disagree, two human annotators inspect
the patch and adjudicate a final label~\cite{ahmed2025can}. 
Next, we describe how we refine the catalog and measure labeling error to contextualize the expected uncertainty. 

\subsubsection{Iterative Refinement}
\label{sec:methodology-pattern-refinement}
The SysLLMatic catalog was only applied to classify benchmark optimizations (\eg HumanEval~\cite{chen2021evaluatinglargelanguagemodels}). 
It had not been validated on real-world PRs, where code changes are larger and noisier.
In an initial pass, LLM agreement was modest, with $\kappa=0.53/0.48$ for agent PRs and $\kappa=0.42/0.35$ for human PRs (category/pattern).
Manual review revealed optimization strategies absent from the original catalog; thus, we refined it using PRs where the two models disagreed.
Two authors (both Ph.D students) independently inspected these cases and revised the catalog iteratively until additional reviews produced no new patterns~\cite{glaser1967discovery}.
This process expanded the catalog from 43 to 59 patterns through 17 additions, 1 deletion, and 5 edits, and introduced two new high-level categories: \textit{Build, Compilation, and Infrastructure Optimization} and \textit{Network, Database, and Data Access Optimization} (see artifact).

With the revised catalog, LLMs agreement improved to $\kappa=0.61/0.47$ for human PRs and $\kappa=0.62/0.47$ for agent PRs (category/pattern).
Remaining disagreement cases were manually labeled by two annotators, with an additional \texttt{No Meaningful Performance Change} label for trivial or non-\texttt{perf} edits. 
Inter-annotator agreement was $\kappa=0.66/0.58$ for human PRs and $\kappa=0.60/0.60$ for agent PRs.
Remaining disagreements were resolved through discussion to produce final consensus labels.
\textit{We additionally measure error rate to assess reliability~\cite{tornberg2024best}.} 
In LLM disagreement cases, alignment with final manual labels is limited (GPT: 35.9\%/18.0\%; Gemini: 38.0\%/22.1\% at category/pattern), improving to 48.3\%/24.1\% and 51.0\%/29.7\% after excluding 50 \texttt{No Meaningful Performance Change} cases.
In LLM-agreed cases, manual auditing of a 10\% random sample (20 PRs) indicates an error rate of 10\%.

\subsection{Testing and Evaluation Behavior (RQ2)}
This question examines the practices used to measure performance (\eg benchmarks, profiling) and how frequently such tests are reported. Because the AIDev dataset does not provide activity-level developer traces, our analysis captures how validation is \textit{described} in pull request artifacts (\ie self-report).\footnote{Observing actual behavior would require fine-grained event data such as an IDE-instrumented dataset (\eg the MSR 2018 Mining Challenge~\cite{msr18challenge}).} 
We use claims of testing and evaluation as evidence that these activities were performed, but acknowledge that LLMs may not be trustworthy in this regard.

We analyze how performance validation is presented by examining
PR descriptions, comments, and documentation, and classify such evidence using the three categories defined in~\cref{sec:background-validation}:
\emph{Benchmark-Based Validation}, \emph{Profiling-Based Validation}, and \emph{Static-Reasoning-Based Validation}.
To account for validation claims that lack quantitative and analytical support, we introduce a fourth category, \emph{Anecdotal or Informal Local Testing}.
PRs with no validation evidence are labeled as having no performance validation.
Following RQ1, we exclude 50 PRs labeled as \texttt{No Meaningful Performance Change}, as validation claims are not meaningful for those PRs.

We use the same LLMs as in RQ1 to independently classify each PR, achieving agreement of $\kappa=0.74$ for validation presence and $\kappa=0.75$ for validation type. For the remaining PRs with model disagreement, two authors independently reviewed and adjudicated labels, with inter-annotator agreement of $\kappa=0.62$ and $\kappa=0.58$, respectively.
\textit{We additionally measure error rates to assess reliability.} 
In LLM disagreement cases, GPT aligns with the manual labels in 32.8\% of validation presence and 12.1\% of validation type, whereas Gemini aligns in 75.9\% and 65.5\%, respectively.
In LLM-agreed cases, we manually audited 10\% of cases (29 PRs).
We observed error rates of 6.9\% for validation presence and 10\% for validation type.

\section{Results}
In this section, we first present a quantitative characterization of the dataset and then answer our two research questions.
Some additional data visualizations are provided in the artifact (\cref{sec:DataAvailability}).
\subsection{Initial Dataset Characterization}

\begin{figure}[h]
    \centering
    \includegraphics[width=\linewidth]{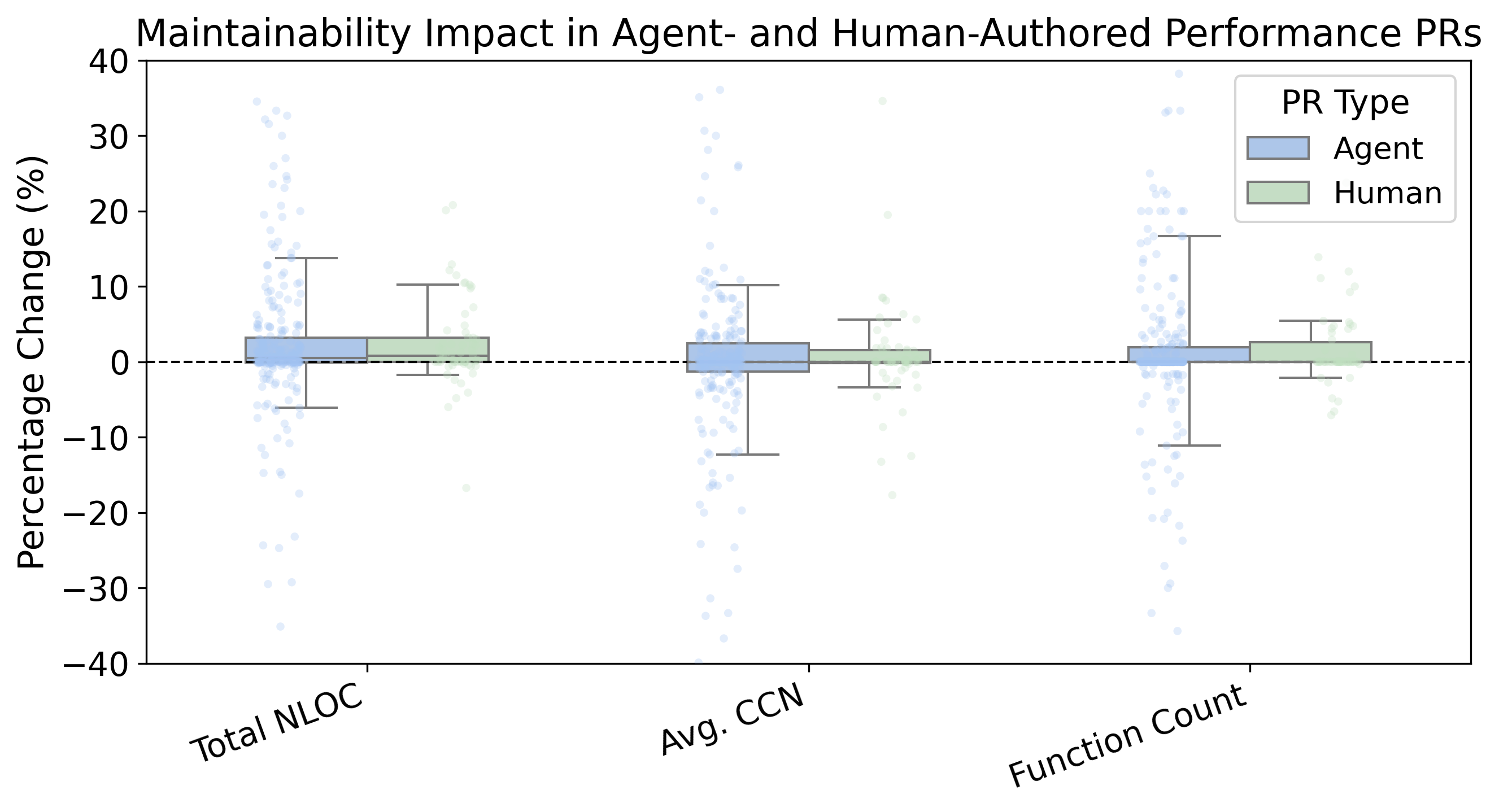}
    \caption{
    Patch size and maintainability measured with Lizard~\cite{lizard_terryyin_2025}. 
    Boxes show medians and interquartile ranges, with whiskers extending to the 10th and 90th percentiles.
    AI-authored PRs show broadly similar medians to human ones, but exhibit higher dispersion and a larger number of extreme outliers.
    }
    \vspace{-1.5em}
    \label{fig:maintaiinability}
\end{figure}

We begin by quantifying AI- and human-authored \texttt{perf} PRs along adoption outcomes (merge rate and time to merge), patch size, and maintainability. 
The result shows that agent \texttt{perf} PRs are merged less frequently than human ones (57\% vs.\ 65\%). 
However, when merged, agent PRs are integrated substantially faster, with a median time to merge of 0.03 hours, compared to 2.65 hours for human ones.
Agent PRs cluster near immediate merge, while human PRs show a broader, long-tailed distribution (Mann-Whitney U~\cite{nachar2008mann}, $p < 0.001$).
In terms of maintainability (Figure~\ref{fig:maintaiinability}), AI- and human-authored PRs show substantial overlap across all three metrics (Total NLOC, AvgCCN~\cite{watson1996structured}, and Function Count), with no significant differences in their central distributions ($p > 0.05$). 
This indicates that the maintainability impact of agent PRs is comparable to that of human-authored patches.
Similar fractions of AI and human PRs increase AvgCCN (40.14\% vs.\ 41.94\%), but agents show a heavier positive tail, indicating larger complexity increases when they occur. 

\subsection{Optimization Patch Characteristics (RQ1)}
\label{sec:result-pattern}

\begin{figure}[h]
    \centering
    \includegraphics[width=\linewidth]{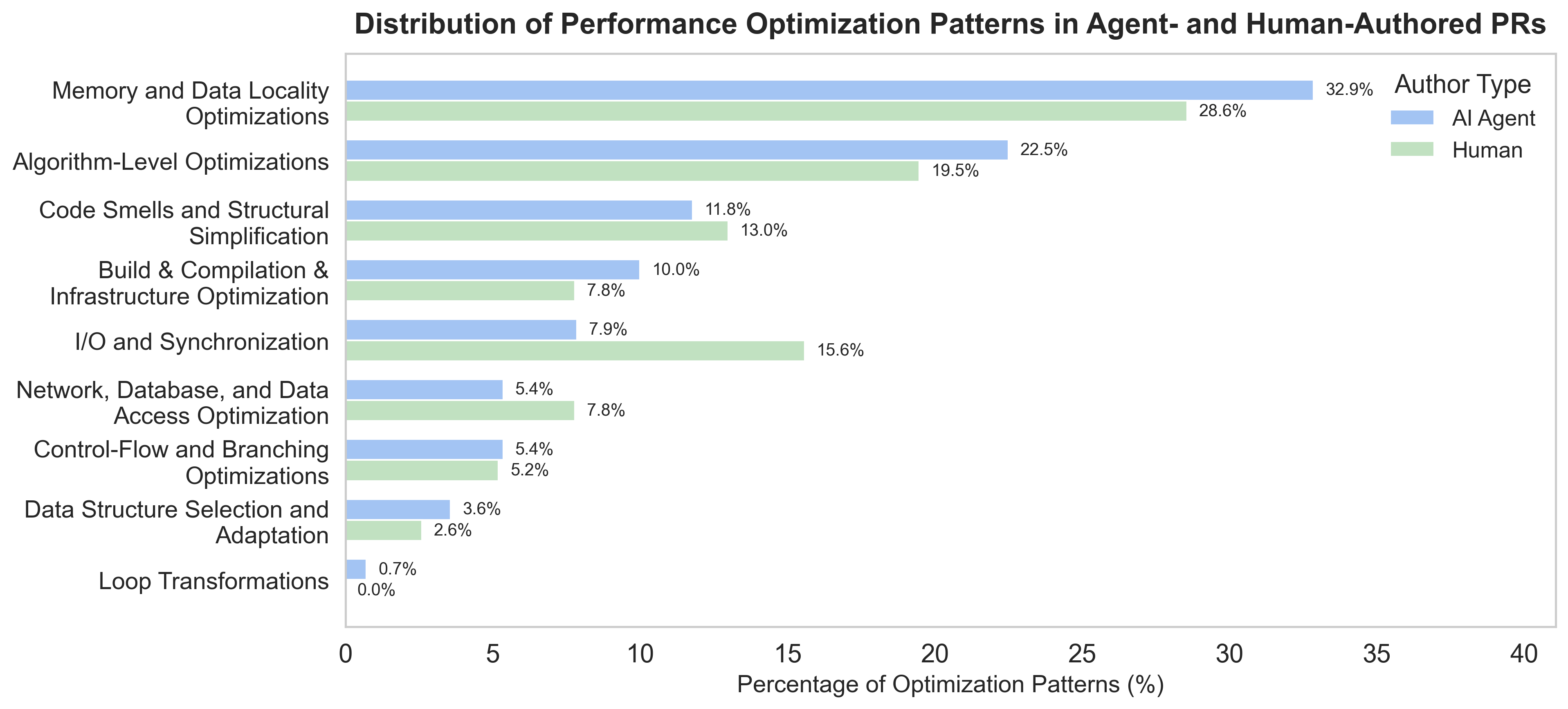}
    \caption{
    Distribution of high-level optimization patterns in AI- and human-authored \texttt{perf} PRs. Both groups exhibit similar distributions, with memory and data-locality optimizations and algorithm optimizations dominating.
    }
    \vspace{-0.5em}
\label{fig:pattern_distribution}
\end{figure}

We analyze the optimization patterns applied by agents and humans across performance patches.
Figure~\ref{fig:pattern_distribution} compares the normalized distributions of high-level optimization patterns between agent and human PRs. 
It shows that the distributions are visually similar, with both groups emphasizing memory/data-locality and algorithm-level optimizations.
Because both variables are categorical, we perform a chi-square test of independence~\cite{agresti2018categorical, mchugh2013chi} on a $2\times9$ contingency table.
The test finds no significant association between author type and optimization category ($\chi^2(8)=6.10$, $p=0.636$; Cramér’s $V=0.13$), indicating that \ul{agents and humans emphasize similar classes of high-level performance optimizations}.



At a finer granularity, agent PRs exhibit a larger set of distinct optimization sub-patterns than human ones (38 vs.\ 21 of 59 total patterns); however, this difference reflects the larger volume of agent PRs rather than systematically broader optimization behavior.
To assess this, we measure sub-pattern richness and control for sample-size imbalance using a label-shuffling permutation test~\cite{good2013permutation} and a subsampling-based rarefaction analysis~\cite{sanders1968marine}.
Neither test finds a statistically significant difference between agent and human PRs (permutation $p=0.12$; rarefaction $p=0.13$), indicating that \ul{agents do not exhibit a wider variety of performance optimization strategies than humans; rather, they apply the same classes of optimizations observed in human-authored PRs}.

\subsection{Testing and Evaluation Behavior (RQ2)}
\label{sec:result_testing}

We analyze whether \texttt{perf} PRs differ by author type in
  (i) the presence of validation,
  and
  (ii) the type of validation employed.


\textit{Validation Presence.}
Among agent PRs, 128 of 280 (45.7\%) include validation, while 49 of 77 (63.6\%) human-authored PRs include validation. A chi-square test of independence indicates a statistically significant but weak association between author type and validation presence ($\chi^2(1)=7.06$, $p=0.007$), with a Cramér’s $V$ of $0.14$~\cite{cohen2013statistical}.

\begin{figure}[h]
    \centering \includegraphics[width=\columnwidth]{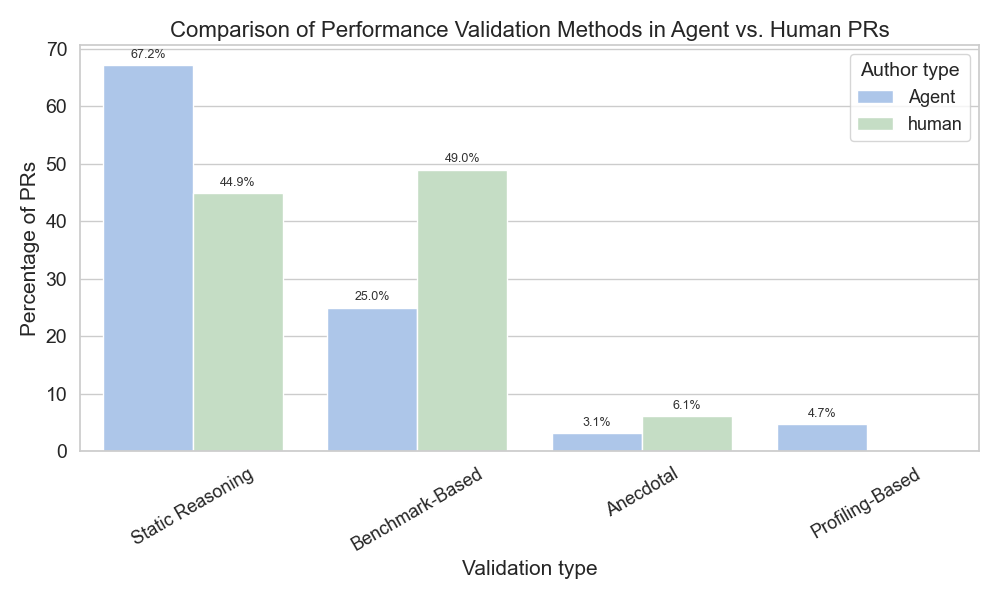}
    \caption{
    Distribution of types of validation reported in \texttt{perf} PRs. Agent PRs rely predominantly on static reasoning checks, whereas human PRs more frequently report benchmark-based validation in addition to static reasoning.
    }
    \vspace{-1.5em}
    \label{fig:validation_type_by_author}
\end{figure}

\textit{Validation Type.}
We further analyze the distribution of validation methods, as shown in~\cref{fig:validation_type_by_author}. Agent-authored PRs predominantly rely on static-reasoning-based validation, which appears in 67.2\% of validated AI PRs, compared to 44.9\% for human PRs. In contrast, humans more frequently report benchmark-based validation, with 49\% of validated human PRs including benchmark results, versus 25\% for agents. 
A chi-square test indicates a statistically significant association between author type and validation method ($\chi^2(3)=12.43$, $p=0.006$), with a Cramér’s $V$ of $0.26
$.

These results indicate that \ul{agents report validation less frequently overall and rely more heavily on static reasoning, whereas humans more frequently use benchmark-based validation when reporting performance improvements}.
We then inspect individual PRs to illustrate the strengths and limitations of agentic validation practices.
We observe that agents can add new benchmarks (\eg inserting \texttt{time.perf\_counter()} around the optimized call) and providing static reasoning based on algorithmic complexity (\eg replacing string concatenation with \texttt{StringBuilder} to reduce complexity from $O(n^2)$ to amortized $O(n)$). However, they may also report benchmark data without supporting evidence (\eg \texttt{7400$\times$ speedup} stated in the PR description without corresponding benchmark code), exposing such claims to the risk of hallucination.



\section{Threats to Validity}
\label{sec:Threats}

We discuss construct, internal, and external threats to validity~\cite{Wohlin2012}.

\underline{Construct}:
We inspect PR artifacts for validation evidence; validation conducted outside the PR workflow or documented informally may lead us to underestimate performance validation practices.
We use LLM-assisted annotation to classify optimization patterns and detect validation evidence in PRs, which may be impacted by hallucinations or misinterpretation of code. We mitigate this by using temperature $=0$, two independent LLMs, and Cohen's $\kappa$ for inter-model agreement; disagreements are resolved by two human annotators, though some subjectivity may remain.

\underline{Internal}:
We use chi-square, permutation, and rarefaction tests. Chi-square assumes independent observations and adequate cell counts, which may be violated by repository/author clustering and sparse categories.
Permutation and rarefaction mitigate sample-size imbalance but assume label exchangeability.
\ifEXTENDED
\else
Significant
\fi
results with small effect sizes should be interpreted as associative, not causal.

\underline{External}:
Our study builds on the AIDev dataset and inherits its limitations, including potential PR author-labeling errors 
and its asymmetric repository filters for agent vs.\ human PRs ($>100$ vs.\ $>500$ stars), which can bias the sample composition and size.

\section{Discussion}

\subsection{Towards Next-Gen Code Optimization}
Our analysis shows that agent PRs draw from a similar optimization pattern space as human ones (\cref{sec:result-pattern}). This alignment highlights both the effectiveness of current LLMs in acquiring established performance engineering knowledge and the limits of their present behavior in practice. 
Despite decades of work showing the effectiveness of loop techniques such as unrolling~\cite{loop_unrolling} and fusion~\cite{loop_fusion}, they remain the least frequently applied category: they are absent from human \texttt{perf} PRs and appear in only 0.7\% of agent-authored ones (\cref{fig:pattern_distribution}). 
In human development workflows, loop techniques are often avoided because they can reduce code readability and maintainability, and are error-prone to implement correctly~\cite{McKinley1996, Mens2004}.

These constraints do not apply in the same way to AI agents. 
As software development becomes increasingly agent-centric, this gap highlights an underexplored space: enabling agents to consider optimization techniques that are traditionally underutilized by human. 
Realizing this potential requires agents tightly integrated with profiling feedback, correctness checks, and automated evaluation pipelines, allowing them to explore a broader optimization space while maintaining the reliability expected of production systems.

\subsection{Performance Validation Gaps in Agent PRs}
Estimating performance impact is challenging --- sensitivity to workloads, hardware, and measurement noise often produces misleading or irreproducible results~\cite{Mytkowicz2009, Papadopoulos2021}. Thus, rigorous improvements require validation through benchmarking or profiling~\cite{ Beyer2019, Hoefler2015}.
Despite this, we found that agent \texttt{perf} PRs include explicit testing or validation substantially less frequently than human ones (\cref{sec:result_testing}). 
In the absence of explicit requirements or safeguards, this gap increases the risk that agent-generated performance changes are integrated without sufficient empirical evidence. 
We further observe a difference in validation style: agents rely on static-reasoning-based justification, whereas humans more often provide benchmark-based evidence (\cref{fig:validation_type_by_author}). 
Although static reasoning can indicate plausible performance effects, it lacks quantitative grounding and is therefore susceptible to incorrect assumptions or hallucinated claims. 

As agent-authored \texttt{perf} PRs grow, future work should consider how agents access and invoke performance evaluation infrastructure, since existing CI-based benchmarking and testing resources, designed around human contribution rates, are unlikely to scale to agent-driven workloads. 
Addressing this mismatch requires standardized benchmarking services and shared profiling resources, and community-level platforms for automated performance validation.

\paragraph*{Data Availability}
\label{sec:DataAvailability}
Our script, data, and prompt are available at: \\ \url{https://anonymous.4open.science/r/perf_patch_study}.

\bibliographystyle{ACM-Reference-Format}
\bibliography{references/reference}

\end{document}